 \documentclass[aps,prb,twocolumn,groupedaddress,showpacs]{revtex4}

\usepackage{graphicx}

\begin{document}

\title{Sigma-phase in the Fe-Re alloy system: experimental and theoretical studies }

\author{J. Cieslak}
\email[Corresponding author: ]{cieslak@fis.agh.edu.pl}
\author{S. M. Dubiel}
\author{J. Zukrowski}
\author{J. Tobola}
\affiliation{AGH University of Science and Technology,
             Faculty of Physics and Applied Computer Science,
             al. Mickiewicza 30, 30-059 Krakow, Poland}

\date{\today}

\begin{abstract}
X-ray diffraction (XRD) and M\"ossbauer spectroscopy techniques combined with theoretical
calculations based on the Korringa-Kohn-Rostoker (KKR) electronic structure calculation
method were used to investigate $\sigma$-phase
Fe$_{100-x}$Re$_{x}$ alloys ($x =$ 43, 45, 47, 49 and 53). Structural data such as site occupancies and
lattice constants were derived from the XRD patters, while the average isomer shift and  distribution
curves of the quadrupole splitting were obtained from the M\"ossbauer spectra. Fe-site
charge-densities and the quadrupole splittings were computed with the KKR method for each lattice site.
The calculated quantities combined with the experimentally determined site occupancies were
successfully used to decompose the measured M\"ossbauer spectra into five components
corresponding to the five sublattices.
\end{abstract}

\pacs{
33.45.+x,       
61.43.-j,       
71.20.Be,       
71.23.-k,       
74.20.Pq,       
75.50.Bb,       
76.80.+y        
}

\maketitle

\section{Introduction}

The $\sigma$-FeRe is one among about fifty members of this phase known to exist in binary alloy
systems, AB, where (larger) A is a transition metal element which belongs to the V-th or VI-th Group
and (smaller) B-one to the VII-th or VIII-th Group of Periodic Table of Elements \cite{Hall66, Sinha72}.
It is also one of five Fe-containing binary $\sigma$-phases.
Its  occurence in the Fe-Re alloy system was definitely established in 1956 \cite{Niemiec56}:
the $\sigma$-phase was obtained by sintering elemental iron and rhenium powders, mixed in the
proportion 3:2, at 1673 K for 4 hours. The relative concentration of the constituting elements was
chosen following a phase diagram by Eggers in which a phase Fe$_3$Re$_2$ was indicated \cite{Eggers38}.
Its characteristic features viz. hardness and brittlenes were similiar to those known for $\sigma$-phases.
Aspecially hard was the Fe$_3$Re$_2$ in the range of 45-50 at\%Re.
Noteworthy, neither systematic studies of $\sigma$ in the Fe-Re alloys aimed at establishing borders of its
existence nor at revealing its physical properties were carried out so far. Among known papers on
the issue, one has to mention the ones reporting a successful synthesis of $\sigma$ by a isothermal annealing
of ingots of Fe$_{55}$Re$_{45}$ alloys at 1763 K for 6 hours \cite{Kopecki59, Ageev60} as
that of Fe$_{53.6}$Re$_{64.4}$ at 1603 K for 8 hours \cite{Joubert08}.

\begin{figure}[b]
\includegraphics[width=.49\textwidth]{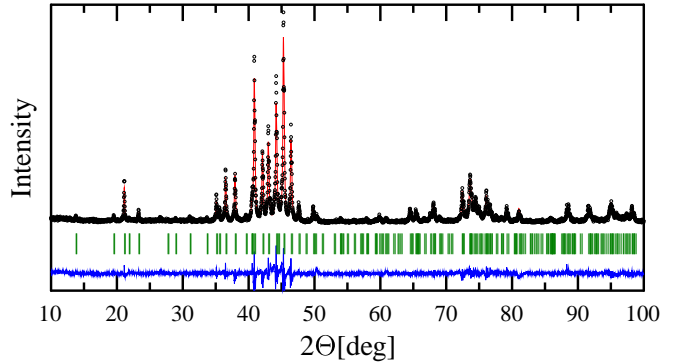}
\caption{(Online color)
Parts of of selected fitted X-ray diffractograms recorded at 294K on the $\sigma$-phase sample of
Fe$_{53}$Re$_{47}$.
The solid line stays for the best-fit obtained with the procedure described in the text.
Peak positions for $\sigma$-phase are indicated, a difference diffractogram is shown, too.
}
\label{fig1}
\end{figure}

Here we report on a successful synthesis of seven samples of $\sigma$-Fe$_{100-x}$Re$_{x}$ alloys with
$x =$ 41, 43, 45, 47, 49, 53 and 55, nominally, and on their experimental and theoretical studies with X-ray diffraction
and M\"ossbauer spectroscopy techniques, as well as electronic structure calculations performed with the
charge and spin self-consistent Korringa-Kohn-Rostoker (KKR) Green's function method \cite{mrs,cpa,stopa}.

\section{Experimental}

\begin{figure}[t]
\includegraphics[width=.49\textwidth]{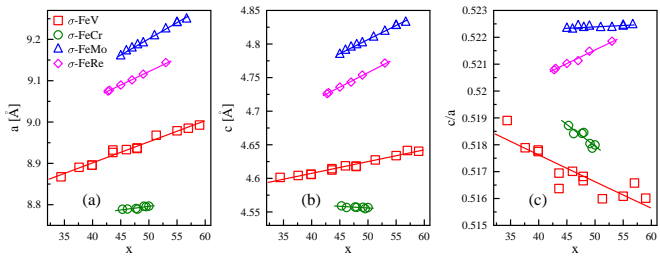}
\caption{(Online color)
Dependence of the lattice parameters $a$  and $c$ as well as $c/a$ ratio for $\sigma$-Fe$_{100-x}$X$_{x}$
(X=Cr, V\cite{Cieslak08c}, Mo\cite{Cieslak12c} and Re)
versus $x$, as determined from the X-ray diffractograms recorded at 294K.
}
\label{fig2}
\end{figure}

The $\sigma$-phase was obtained in the following way: powders of elemental iron (99.9+ purity) and
rhenium (99.99 purity) were mixed in appropriate proportions and mases of $\sim 2$g were next pressed
to pallets. The pallets were subsequently  melted in an arc furnace under protective
atmosphere of argon. The ingots were next re-melted three times to improve their
homogeneity. Finally, they were vacuum annealed at 1800 K for 5 hours and quenched into
liquid nitrogen. The mass loses of the fabricated $\sigma$-FeRe alloys were no more than 0.01\%
of their initial values, so it is reasonable to take their nominal compositions as real ones.
The samples were investigated with two experimental techniques viz. X-ray diffraction (XRD)
and the M\"ossbauer spectroscopy (MS). Measurements of diffraction patters and of M\"ossbauer
spectra were carried out on powdered samples at room temperature (the $\sigma$-phase is very
brittle, so it could be easily transformed into powder by attrition in an agate mortar). From the
XRD patters, an example of which is displayed in Fig. 1, we obtained an evidence that in all
cases, except two border compositions, the transformation into the $\sigma$-phase was 100\% successful.
In the sample with the highest Re-content ($x =$ 55) the transformation was not fully complete, so this
sample was excluded from further investigations. On the other hand, the sample with the highest
Fe-content ($x =$ 41) was found to be fully transforemed and no any traces of other phases were found.
Unfortunatelly, the values of lattice constants as well as the volume of the unit cell did not stay
in line with the corresponding results for other samples, namely they are shifted to $x =$ 42.8 so to higher Re concentrations.
These values should be taken as borders of the formation of the FeRe $\sigma$-phase at 1800 K.

\begin{figure}[b]
\includegraphics[width=.49\textwidth]{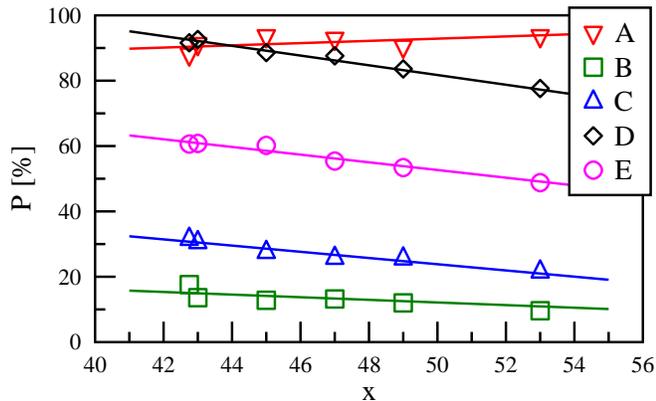}
\caption{(Online color)
Probability of finding Fe atoms at different lattice sites in the
$\sigma$-Fe$_{100-x}$Re$_x$ compounds, $P$,  versus Re concentration, $x$. Solid lines
stay for the linear fits to the data}
\label{fig3}
\end{figure}

\section{Results and discussion}

\subsection{XRD measurements}

\begin{table} 
\caption{\label{table1} Atomic crystallographic positions for the five
lattice sites of the Fe-Re $\sigma$-phase.}
\begin{tabular}{|l|l|l|l|l|} \hline
Site&Wyckoff index    & x          & y             & z             \\ \hline
A   & 2i & 0          & 0             & 0             \\ \hline
B   & 4f & 0.4016(3)  & 0.4016(3)     & 0             \\ \hline
C   & 8i & 0.4645(2)  & 0.1327(4)     & 0             \\ \hline
D   & 8i'& 0.7422(5)  & 0.0657(3)     & 0             \\ \hline
E   & 8j & 0.1834(2)  & 0.1834(2)     & 0.2497(3)     \\ \hline
\end{tabular}
\end{table}

\begin{table}[t] 
\caption{\label{table2} Lattice constants $a$ and $c$ as measured for all investigated
$\sigma$-Fe$_{100-x}$Re$_x$ samples.
}
\begin{tabular}{|l|l|l|} \hline
$x$ & $a$ [\AA]         & $c$ [\AA]                       \\ \hline
42.8& 9.0749(2)         & 4.7261(1)                       \\ \hline
43.0& 9.0767(1)         & 4.7276(1)                       \\ \hline
45.0& 9.0895(2)         & 4.7359(1)                       \\ \hline
47.0& 9.1019(1)         & 4.7433(1)                       \\ \hline
49.0& 9.1159(1)         & 4.7538(1)                       \\ \hline
53.0& 9.1438(1)         & 4.7717(1)                       \\ \hline
\end{tabular}
\end{table}

The powder XRD patterns were collected at RT with a D5000
Siemens diffractometer (using Cu K-$\alpha$ radiation and a graphite secondary
monochromator) within the 2$\theta$-rande from 10$^\circ$ to 140$^\circ$ in steps of 0.02$^\circ$.
Data were analyzed by the Rietveld method as implemented in the FULLPROF program
\cite{Rodriguez93} to get information on the
crystallographic structure and the sites occupancy. From 22 free parameters used, 6 were
related to a background and line positions, 10 to lattice sites occupancies and atomic positions,
while the remaining 6 to line widths, lattice constants and Debye-Waller factors. The analysis
yielded the lattice constants $a$ and $c$  (Table \ref{table2}, Fig. 2a,b), the atomic positions (Table \ref{table1}), and
the lattice sites occupancies (Fig. 3).

\begin{table*} 
\caption{\label{table3} Fe-occupancies, isomer shift, $IS$ and quadrupole splitting, $QS$ for the $\sigma$-Fe$_{53}$X$_{47}$
(X=Cr\cite{Cieslak08c,Cieslak08b}, V\cite{Cieslak08c,Cieslak10a}, Mo\cite{Cieslak12c} and Re)}

\begin{tabular}{|l|l|l|l|l|l|l|l|l|l|l|l|l|} \hline
    &\multicolumn{4}{c|}{Fe-occupancy [\%]}&  \multicolumn{4}{c|}{IS [mm/s]}&   \multicolumn{4}{c|}{QS [mm/s]}     \\ \hline
Site&   V   &  Cr &   Mo  &   Re  &       V&     Cr&    Mo &    Re &        V&     Cr&    Mo &    Re          \\ \hline
A   &  93.2 & 87.8& 100.0 &  91.8 &      - &    -  &   -   &   -   &    0.351& 0.342 & 0.280 & 0.598          \\ \hline
B   &  18.2 & 27.2&  12.9 &  13.3 &   0.341& 0.351 & 0.303 & 0.260 &    0.292& 0.242 & 0.404 & 0.616          \\ \hline
C   &  35.2 & 40.5&  25.5 &  26.7 &   0.201& 0.216 & 0.205 & 0.246 &    0.282& 0.181 & 0.329 & 0.587          \\ \hline
D   &  96.3 & 89.2& 100.0 &  86.2 &   0.012& 0.023 & 0.023 & 0.067 &    0.209& 0.210 & 0.208 & 0.352          \\ \hline
E   &  34.9 & 33.5&  41.9 &  56.2 &   0.115& 0.113 & 0.113 & 0.141 &    0.454& 0.454 & 0.464 & 0.640          \\ \hline
\end{tabular}
\end{table*}

Concerning the atomic positions, they do not depend on the samples composition
within the error limit. They are also in line with those reported in the literature
\cite{Joubert08}. Both lattice constants show a linear dependence on the composition viz.
they increase with $x$. Such behavior is consistent with that found for the
$\sigma$-phase in other binary Fe-based alloys, namely Fe-Cr, Fe-V and Fe-Mo, and it is
related to the atomic size of the so-called A element (here Cr, V, Mo, Re) which is
larger than the size of Fe atom (1.56 \AA). Also relative values of the lattice constants
reflect the atomic size effect as their relative ordering follows the atomic size of Cr
(1.66 \AA), V(1.71 \AA), Re (1.88 \AA) and Mo (1.90 \AA).  A different trend, as illustrated in Fig. 2c,
can be seen for the $c/a$ ratio which is for the $\sigma$-FeRe concentration dependent
increasing from 0.5207(1) to 0.5219(1). These values can be compared with 0.5211(1) given by Joubert
\cite{Joubert08}. For all other $\sigma$-phase compounds shown in this figure, the ratio either
stay constant (Fe-Mo) or linearly decreases with the concentration
of the element A (Cr, V).

\begin{figure}[t]
\includegraphics[width=.50\textwidth]{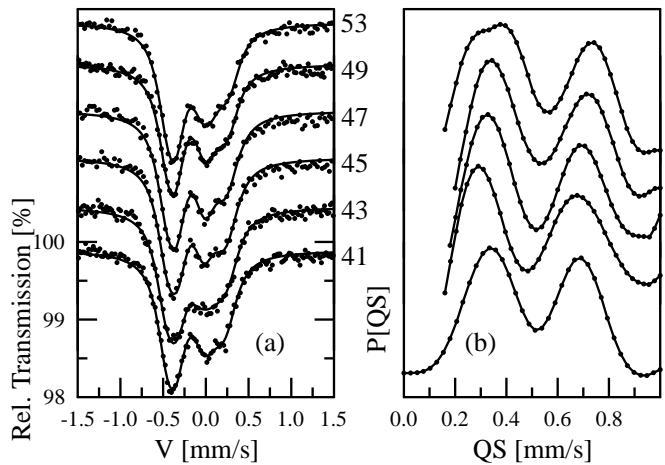}
\caption{
(a) $^{57}$Fe M\"ossbauer spectra recorded on a series of $\sigma$-Fe$_{100-x}$Re$_x$ samples
at 294K and marked with the corresponding $x$-values. The solid lines
are the best-fit to the experimental data. The derived quadrupole splitting distribution curves
shown in the same sequence as the spectra, are indicated in (b).
}
\label{fig4}
\end{figure}

Regarding the sites occupancies, all five sites are mixed i.e. occupied by both types of
atoms - see Fig. 3 and Table \ref{table3}. However, sites A and D are in majority populated by
Fe atoms,  while sites B and C are mostly occupied by Re atoms. The population of
Fe/Re atoms on site E is rather in balance, yet the probability of finding an Fe atom
on this site linearly decreases with $x$ from $\sim 60$\% for $x =$ 43 to $\sim 50$\% for $x =$
53. A similar trend is observed for the sites B, C and D. An exceptional trend exhibits
the site A for which the occupancy in Fe atoms slightly increase with $x$.

It is interesting to compare the sites occupancies in different Fe-X alloy systems. A
range of composition where the $\sigma$-phase can be formed is characteristic of the
system. However, the range of $x = 45-50$ is common, so a comparison of the sites
occupancies for a concentration from that range seems reasonable. Appropriate data
for $x =$ 47 are displayed in Table \ref{table3}. It is evident that the sites A and D are
predominantly (exclusively, for Fe-Mo) occupied by Fe atoms irrespective of the
alloy system. The lowest population of Fe atoms is typical of the site B (27\%, at
maximum). The occupancies of the sites C and E behave in a way characteristic of the
A element: for Cr the site C is more populated by Fe atoms than the site E, while for
Mo and Re the opposite is true. Finally, for V the C and E sites host about the same
number of  Fe atoms.

\begin{figure}[t]
\includegraphics[width=.49\textwidth]{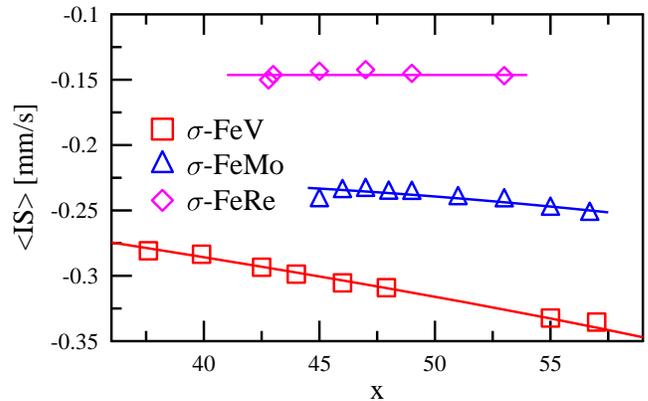}
\caption{(Online color)
The average isomer shift (relative to the Co/Rh source), $\langle IS \rangle$, versus concentration, $x$, as measured (markers)
and as calculated (lines) for $\sigma$-Fe$_{100-x}$X$_{x}$
(X=Cr\cite{Cieslak08b}, V\cite{Cieslak10a}, Mo\cite{Cieslak12c} and Re).
}
\label{fig5}
\end{figure}

\begin{figure}[b]
\includegraphics[width=.49\textwidth]{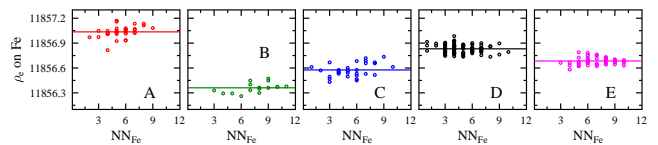}
\caption{(Online color)
Fe-site charge density, $\rho_e$, for five crystallographic sites versus $NN_{Fe}$.
Solid lines represent average values of $\rho_e$}
\label{fig6}
\end{figure}

\subsection{M\"ossbauer measurements}

$^{57}$Fe spectra which were recorded at 295 K in a transmission mode using a standard
spectrometer and a $^{57}$Co/Rh source for the gamma rays are presented in Fig. 4. They are very
asymmetric, which was not observed for other $\sigma$-phase Fe-X (X=Cr, V, Mo) alloys. But
as in the previous cases, the structure of the spectra is badly resolved which did not allow to
analyze them in terms of a superposition of subspectra corresponding to the five sublattices.
Instead, they were fitted assuming a distribution of the quadrupole splitting, $QS$, and a
linear correlation between  $QS$ and the isomer shift, $IS$ \cite{Caer98}. The analysis yielded
the distribution curves of $QS$, that are displayed in Fig. 4b, as well as the average isomer
shifts, $\langle IS \rangle$ - see Fig. 5. Concerning the $QS$-values, it is evident that their distribution has
two maxima: one at about 0.35 mm/s, the other at about 0.75 mm/s. The average isomer shift
does not practically depend on the composition which means the average Fe-site charge
density is the same in all investigated samples. This is different in comparison with the $\sigma$-
phase in Fe-Mo and Fe-V alloys, where, as indicated in Fig. 5, one observes a weak linear
decrease of $\langle IS \rangle$ with the concentration of Mo and V, respectively \cite{Cieslak10a,
Cieslak12c}.

\subsection{Electronic structure calculations}

\begin{figure}[t]
\includegraphics[width=.49\textwidth]{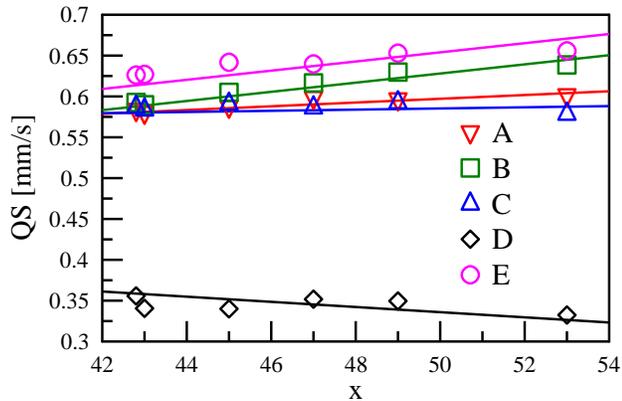}
\caption{(Online color)
Quadrupole splitting, $QS$, as determined
for each site and Re concentration, $x$, from the analysis of
the measured spectra with the protocol described in Ref. \onlinecite{Cieslak08b}.
}
\label{fig7}
\end{figure}

The electronic structure of the $\sigma$-phase can be calculated at the level of particular lattice
sites as already shown for the $\sigma$-phase in  Fe-X (X=Cr, V, Mo) alloys \cite{Cieslak08b,
Cieslak10a, Cieslak12c}. In particular, one can calculate $QS$ and $IS$- values for each site,
and next use them to successfully fit the M\"ossbauer spectra. In the present case, the
calculations were carried out for 16 unit cells having different configurations of Fe/Re atoms over
the five sublattices. The configurations were chosen following two criteria: (1) the probability
of finding Fe/Re atoms on a given sublattice should be as close as possible to the one
determined experimentally, and (2) each possible configuration of atoms with a given number
of Fe atoms occupying the nearest-neighbour position, $NN$,  should be represented at least
once. Charge-densities obtained with this protocol, $\rho _{Fe}$, were expressed in terms of the number
of Fe atoms situated in $NN$, $NN_{Fe}$, for each of the five sublattices. Using next a linear
relationship between IS and $\rho_{e}$ \cite{Neese05}, the calculated $\rho_{e}$-values for each sublattice were re-
calculated into $IS$-values, and finally into the average one for each spectrum, $\langle IS \rangle$. The
calculated potentials in combination with the experimentally determined site occupancies were
used to determine the $QS$-values for each sublattice and composition. The calculations were
done using the extended point-charge model as outlined elsewhere \cite{Cieslak08b, Cieslak10a}.
The charge-densities obtained with the above mentioned protocol and method are visualized in
Fig. 6 for each sublattice. As in the case of $\sigma$-FeMo \cite{Cieslak12c}, the $\rho_{e}$-values are rather
$NN_{Fe}$ independent which was not the case for $\sigma$ in Fe-Cr and Fe-V alloys \cite{Cieslak08b,
Cieslak10a}.  Consequently, the average values of the charge-densities, hence those of the isomer
shifts for the given lattice sites were calculated as an average over all atomic
configurations taken into account in the calculations for that sites. The $\langle IS \rangle$-values were calculated
as the average over all five sites and were in a good agreement with those found from the M\"ossbauer spectra fitting
(Fig. 5). The data shown in Fig. 6 give a possibility for making a comparison between the
charge-densities on particular lattice sites.  It is evident that the highest charge-density have Fe
atoms on sites A, followed by those on D, E, C and B. One can also make a comparison
between the calculated site densities for the five sublattices in different alloy systems we
studied so far. The appropriate data, expressed in terms of the isomer shifts,  are displayed in
Table \ref{table3}. The $IS$-values of particular sites are given relative to that of the site A which was
found to have the highest charge-density.  It is evident that for all studied alloys that the
smallest $IS$-value, hence the highest charge-density among the remaining four sites, have Fe
atoms on the site D followed by the sites E, C and B.

\begin{figure}[b]
\includegraphics[width=.49\textwidth]{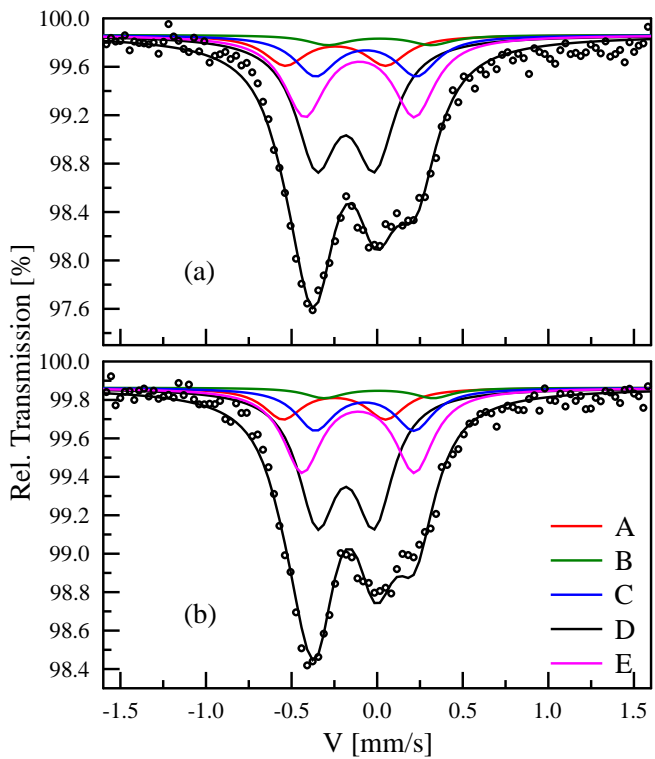}
\caption{(Online color)
$^{57}$Fe-site M\"ossbauer spectra recorded at
294K on two studied samples viz. with (a) $x=45$ and (b) $x=53$. The best-fit
spectrum and five subspectra are indicated by solid lines.}
\label{fig8}
\end{figure}

The calculated values of the quadrupole splitting, $QS$, are shown in Fig. 7. It is clear that the
lowest field gradient was found on the site D, QS = 0.3-0.35 mm/s, while those on the other
four sites are much larger (almost by a factor of 2) and close to each other. The difference
between them slightly increases with $x$. Based on these calculations one can understand the
two-peak structure of the experimentally obtained $QS$-distribution curves (Fig. 4b). A set the
$QS$-values calculated for the different lattice sites and alloys for $x =$ 0.47 is presented in Table
\ref{table3}. It is clear that the $QS$-values found for the Fe-Re samples are meaningfully higher than the
corresponding quantities calculated for the other studied Fe-X alloys. This observation seems
to agree with the fact that Re atoms have a large atomic size (only Mo atoms are larger here,
but the unit cell of the FeRe $\sigma$-phase is smaller), hence they cause the
asymmetric-most distribution of the charge around the probe Fe atoms. Noteworthy, the
smallest $QS$-values were calculated for the site D, and the largest ones for the site E.

\subsection{M\"ossbauer spectra analysis}
The knowledge of the lattice site occupancies combined with that of the calculated $IS$ and $QS$
values permitted to carry out  the analysis of the M\"ossbauer spectra in terms of the five
sublattices. A successful analysis could have been done with only five free parameters. Four
of them viz. background, spectral area, line width (common for all five subspectra) and the
isomer shift of one subspectrum depend on the conditions of the spectra measuremsnts, hence
they could not be calculated. The fifth free parameter was a proportionality constant between the
calculated maximum component of the electric field gradient, $V_{zz}$, in which only the so-called
lattice contribution was taken into account,  and $QS$, in which also a contribution from
electrons localized on the probe atom is included. Two examples of the measured spectra
analyzed in this way are presented in Fig. 8 together with the subspectra belonging to the five
sublattices. A very good agreement between the measured and fitted spectra was achived.

\section{Summary and conclusions}

Six samples of the $\sigma$-phase Fe$_{100-x}$Re$_{x}$ alloys ($x =$ 43, 45, 47, 49, 53 and 55) were synthesized
and experimentally investigated using X-ray diffraction and M\"ossbauer spectroscopy
techniques, and theoretically with the electronic structure calculations using Korringa-Kohn-Rostoker method. XRD
yielded information on site occupancies and lattice constants, MS on the average isomer shift
and a distribution of the quadrupole splitting. Charge-densities and the quadrupole splittings
were computed with the KKR method. The calculated quantities combined with the
experimentally determined site occupancies were successfully used to analyze the measured
M\"ossbauer spectra.

Based on the results reported in this paper, the following conclusions can be made:

\begin{itemize}
\item A pure  $\sigma$-phase in the Fe$_{100-x}$Re$_{x}$ alloy system at 1800K can be formed for $x$ between 43 and 53.

\item Its lattice parameters $a$ and $c$ show a linear increase with $x$, the $c/a$ ratio also increases.

\item The site occupancies are mixed i.e. both Fe and Re atoms can be found on all five sites.
However, the sites A and D are predominantly occupied by Fe atoms, while Re atoms are in
majority on the sites C and B. The population of both types of atoms on the site E is in a fair
balance.

\item The calculated Fe-site charge-density, $\rho_{e}$, are characteristic of the lattice site, they are
rather independent of the number of Fe atoms in the first-neighbor shell, and they
increase in the following order: $\rho_{e}(A) > \rho_{e}(D) > \rho_{e}(E) > \rho_{e}(C) > \rho_{e}(B)$.

\item The average isomer shift, $\langle IS \rangle$, derived from the analysis of the measured spectra is
independent of the alloys composition. This behavior can be explained by the
calculated charge-densities.

\item The calculated quadrupole splitting has the lowest value for the site D, while its values for
the other sites are close to each other and are by a factor of about two
higher. The calculations are in line with the two-peak structure of the $QS$-distribution curves
derived from the measured spectra.

\item The calculated hyperfine quantities permitted to successfully decompose the measured
spectra into five components corresponding to particular lattice sites.
\end{itemize}

\begin{acknowledgments}
This work was supported by the
Ministry of Science and Higher Education, Warsaw (grant No. N N202 228837).

\end{acknowledgments}


\begin{thebibliography}{58}


\expandafter\ifx\csname natexlab\endcsname\relax\def\natexlab#1{#1}\fi
\expandafter\ifx\csname bibnamefont\endcsname\relax
  \def\bibnamefont#1{#1}\fi
\expandafter\ifx\csname bibfnamefont\endcsname\relax
  \def\bibfnamefont#1{#1}\fi
\expandafter\ifx\csname citenamefont\endcsname\relax
  \def\citenamefont#1{#1}\fi
\expandafter\ifx\csname url\endcsname\relax
  \def\url#1{\texttt{#1}}\fi
\expandafter\ifx\csname urlprefix\endcsname\relax\def\urlprefix{URL }\fi
\providecommand{\bibinfo}[2]{#2}
\providecommand{\eprint}[2][]{\url{#2}}


\bibitem{Hall66}
\bibinfo{author}{\bibfnamefont{E.~D.}~\bibnamefont{Hall}} \bibnamefont{and}
  \bibinfo{author}{\bibfnamefont{S.~H.} \bibnamefont{Algie}},
  \bibinfo{journal}{Metall. Rev.} \textbf{\bibinfo{volume}{11}},
  \bibinfo{pages}{61} (\bibinfo{year}{1966}).

\bibitem{Sinha72}
\bibinfo{author}{\bibfnamefont{A.~K.}~\bibnamefont{Sinha}},
  \bibinfo{journal}{Prog. Mater. Sci.} \textbf{\bibinfo{volume}{15}},
  \bibinfo{pages}{79} (\bibinfo{year}{1972}).



\bibitem{Niemiec56}
  \bibinfo{author}{{J. Niemiec}}
  and
  \bibinfo{author}{{W. Trzebiatowski}},
  \bibinfo{journal}{Bull. Acad. Pol. Sci.}
   \textbf{\bibinfo{volume}{4}},
  \bibinfo{pages}{601}
 (\bibinfo{year}{1956})

\bibitem{Eggers38}
  \bibinfo{author}{{H. Eggers}},
  \bibinfo{journal}{Mitt. Kaiser-Wilhelm-Inst., Eisenforsch.}
   \textbf{\bibinfo{volume}{20}},
  \bibinfo{pages}{147}
 (\bibinfo{year}{1938})

\bibitem{Kopecki59}
  \bibinfo{author}{{C.V. Kopetskii}},
  \bibinfo{author}{{V.S. Shekhtman}},
  \bibinfo{author}{{N.V. Ageev}}
  and
  \bibinfo{author}{{E.M. Savitskij}},
  \bibinfo{journal}{Dokl. Akad. Nauk SSSR}
   \textbf{\bibinfo{volume}{125}},
  \bibinfo{pages}{87}
 (\bibinfo{year}{1959})

\bibitem{Ageev60}
  \bibinfo{author}{{N.V. Ageev}}
  and
  \bibinfo{journal}{Dokl. Akad. Nauk SSSR}
   \textbf{\bibinfo{volume}{135}},
  \bibinfo{pages}{309}
 (\bibinfo{year}{1960})

\bibitem{Joubert08}
  \bibinfo{author}{\bibfnamefont{J.-M.}~\bibnamefont{Joubert}}
  \bibinfo{journal}{Progr. Mater. Sci.} \textbf{\bibinfo{volume}{53}},
  \bibinfo{pages}{528} (\bibinfo{year}{2008}).

\bibitem[{\citenamefont{ed. by et~al.}(1992)\citenamefont{ed. by, Butler,
  Dederichs, Gonis, and Weaver}}]{mrs}
  \bibinfo{author}{\bibnamefont{ed. by}}, \bibinfo{author}{\bibfnamefont{W.~H.}
  \bibnamefont{Butler}},
  \bibinfo{author}{\bibfnamefont{P.}~\bibnamefont{Dederichs}},
  \bibinfo{author}{\bibfnamefont{A.}~\bibnamefont{Gonis}}, \bibnamefont{and}
  \bibinfo{author}{\bibfnamefont{R.}~\bibnamefont{Weaver}},
  \emph{\bibinfo{title}{Chapter III, in: Applications of Multiple Scattering
  Theory to Materials Science}}, vol. \bibinfo{volume}{253}
  (\bibinfo{publisher}{MRS Symposia Proceedings, MRS Pittsburgh.},
  \bibinfo{year}{1992}).


\bibitem[{\citenamefont{Stopa et~al.}(2004)\citenamefont{Stopa, Kaprzyk, and
  Tobola}}]{stopa}
  \bibinfo{author}{\bibfnamefont{T.}~\bibnamefont{Stopa}},
  \bibinfo{author}{\bibfnamefont{S.}~\bibnamefont{Kaprzyk}}, \bibnamefont{and}
  \bibinfo{author}{\bibfnamefont{J.}~\bibnamefont{Tobola}},
  \bibinfo{journal}{J. Phys.: Condens. Matter} \textbf{\bibinfo{volume}{16}},
  \bibinfo{pages}{4921} (\bibinfo{year}{2004}).


\bibitem[{\citenamefont{Bansil et~al.}(1999)\citenamefont{Bansil, Kaprzyk,
  Mijnarends, and Tobola}}]{cpa}
  \bibinfo{author}{\bibfnamefont{A.}~\bibnamefont{Bansil}},
  \bibinfo{author}{\bibfnamefont{S.}~\bibnamefont{Kaprzyk}},
  \bibinfo{author}{\bibfnamefont{P.~E.} \bibnamefont{Mijnarends}},
  \bibnamefont{and} \bibinfo{author}{\bibfnamefont{J.}~\bibnamefont{Tobola}},
  \bibinfo{journal}{Phys.\ Rev.\ B} \textbf{\bibinfo{volume}{60}},
  \bibinfo{pages}{13396} (\bibinfo{year}{1999}).

\bibitem{Cieslak08c}
  \bibinfo{author}{\bibfnamefont{J.}~\bibnamefont{Cieslak}},
  \bibinfo{author}{\bibfnamefont{M.}~\bibnamefont{Reissner}},
  \bibinfo{author}{\bibfnamefont{S.~M.}~\bibnamefont{Dubiel}},
  \bibinfo{author}{\bibfnamefont{J.}~\bibnamefont{Wernisch}}
  \bibnamefont{and}
  \bibinfo{author}{\bibfnamefont{W.}~\bibnamefont{Steiner}},
  \bibinfo{journal}{J. Alloys Comp.} \textbf{\bibinfo{volume}{460}},
  \bibinfo{pages}{20} (\bibinfo{year}{2008}).


\bibitem{Cieslak12c}
  \bibinfo{author}{\bibfnamefont{J.}~\bibnamefont{Cieslak}},
  \bibinfo{author}{\bibfnamefont{S.~M.}~\bibnamefont{Dubiel}},
  \bibinfo{author}{\bibfnamefont{J.}~\bibnamefont{Przewoznik}}
  \bibnamefont{and}
  \bibinfo{author}{\bibfnamefont{J.}~\bibnamefont{Tobola}},
  \bibinfo{journal}{To be publisched} \textbf{\bibinfo{volume}{}},
  \bibinfo{pages}{} (\bibinfo{year}{2012}).

\bibitem{Rodriguez93}
  \bibinfo{author}{\bibfnamefont{J. Rodriguez-Carjaval}},
  \bibinfo{journal}{Phisica B} \textbf{\bibinfo{volume}{192}},
  \bibinfo{pages}{55} (\bibinfo{year}{1993}).

\bibitem{Cieslak08b}
  \bibinfo{author}{{J. Cieslak}},
  \bibinfo{author}{{J. Tobola}},
  \bibinfo{author}{{S.~M. Dubiel}},
  \bibinfo{author}{{S. Kaprzyk}},
  \bibinfo{author}{{W. Steiner}}
  and
  \bibinfo{author}{{M. Reissner}},
  \bibinfo{journal}{J. Phys.: Condens. Matter.} \textbf{\bibinfo{volume}{20}},
  \bibinfo{pages}{235234} (\bibinfo{year}{2008}).

\bibitem{Cieslak10a}
  \bibinfo{author}{\bibfnamefont{J.}~\bibnamefont{Cieslak}},
  \bibinfo{author}{\bibfnamefont{J.}~\bibnamefont{Tobola}},
  \bibnamefont{and}
  \bibinfo{author}{\bibfnamefont{S.~M.}~\bibnamefont{Dubiel}},
  \bibinfo{journal}{Phys.\ Rev.\ B} \textbf{\bibinfo{volume}{81}},
  \bibinfo{pages}{174203} (\bibinfo{year}{2010}).

\bibitem{Caer98}
  \bibinfo{author}{\bibfnamefont {G.}~\bibnamefont {Le Caer}},
  and
  \bibinfo{author}{\bibfnamefont {R.~A.}~\bibnamefont{Brandt}},
  \bibinfo{journal}{J. Phys. Condens. Matter}, \textbf{\bibinfo{volume}{10}}
  \bibinfo{pages}{10715} (\bibinfo {year} {1998}).

\bibitem{Neese05}
\bibinfo{author}{\bibfnamefont{F.}~\bibnamefont{Neese}},
  \bibinfo{journal}{Inorganica Chimica Acta} \textbf{\bibinfo{volume}{337}},
  \bibinfo{pages}{181} (\bibinfo{year}{2002}).



\end{thebibliography}
\end{document}